\documentclass[10pt]{article}
\usepackage{graphicx}
\usepackage{fix-cm}
\usepackage{amsfonts}
\usepackage{amssymb}
\usepackage{amsmath}
\newcommand{\bra}[1]{\langle #1|}
\newcommand{\ket}[1]{|#1\rangle}
\newcommand{\braket}[2]{\langle #1|#2\rangle}
\newcommand{\ketbra}[2]{| #1 \rangle \langle #2 |}
\DeclareMathOperator{\Tr}{Tr}
\begin{document}

\title{From Hardy Spaces to Quantum Jumps:\\A Quantum Mechanical Beginning of Time}

\author{Arno Bohm \and Peter W. Bryant }

\maketitle

\begin{center}
              Center for Complex Quantum Systems,
              Department of Physics,\\
              University of Texas at Austin, Austin, Texas 78712
\end{center}

\begin{abstract}
In quantum mechanical experiments one distinguishes between the state of
an experimental system and an observable measured in it.
Heuristically, the distinction between states and observables
is also suggested in scattering theory or when one expresses causality.
We explain how this distinction can be made also mathematically.
The result is a theory with
asymmetric time evolution and for which decaying states are exactly unified with resonances.
A consequence of the asymmetric time evolution is a beginning of time.
The meaning of this beginning of time can be understood by identifying it in
data from quantum jumps experiments.
\end{abstract}

\section{Introduction}
Mathematics is the language of physics;
it is the means to understand and communicate physical results.
But this mathematical language changes as the physical subject develops and expands.
For quantum physics the original mathematical framework was the Hilbert space,
which was
developed on the basis of matrices and wave equations.
As the understanding of quantum physics improved with the analysis of scattering
and decay phenomena, however, the mathematical theory had to be refined.
This led from Dirac kets to Schwartz's distribution and to the 
Rigged Hilbert Space of Gelfand and Maurin,
$\Phi\subset\mathcal{H}\subset\Phi^\times$,
and to the development of a pair of Hardy Rigged Hilbert Spaces.
For a recent review, see~\cite{bohm_review_rmp_2009}.

For a theory that unifies resonance scattering and decay phenomena, one needed a pair
of Rigged Hilbert Spaces to discriminate between the two kinds of Lippmann-Schwinger
kets, as well as between prepared in-states and detected out-observables.
The mathematics for this unified theory is given by Gadella's
Diagrams~\eqref{gd:obs}, \eqref{gd:states},
which define the underlying mathematical theory~\cite{civitarese_physical_2004}.

With the mathematics defined, one can make rigorous predictions.
One of the predictions,
that lifetime is exactly equal to the inverse width, may make some physicists happy.
Another prediction is that the time evolution of a quantum state starts at a finite
time, $t_0: t_0\leq t < \infty$, rather than extending over the range
$-\infty < t < \infty$~\cite{bohm_early_theory_1981}.
The appearance in experiments of this quantum mechanical beginning of time, $t_0$,
will be discussed in the last section of this paper.

\section{Distinguishing States and Observables}
\label{sec:statesobs}
In experiments with quantum systems one distinguishes between states and observables.\\[5pt]
\begin{minipage}{0.40\textwidth}
  State $\rho$ of a system (e.g. \textit{in-}states $\phi^{+}$ of scattering experiment.)
\end{minipage}\hfill
\begin{minipage}{0.32\textwidth}
\begin{center}
\emph{is prepared by}
\end{center}
\end{minipage}\hfill
\begin{minipage}{0.27\textwidth}
preparation apparatus (e.g. accelerator)
\end{minipage}\\[5pt]
\begin{minipage}{0.40\textwidth}
Observable $A$ (e.g. detected \textit{out-}observables $\psi^-$ of
a scattering experiment, ``out-state'')
\end{minipage}\hfill
\begin{minipage}{0.32\textwidth}
\begin{center}
\emph{is registered by}
\end{center}
\end{minipage}\hfill
\begin{minipage}{0.27\textwidth}
registration apparatus (e.g. detector)
\end{minipage}\\[5pt]
\begin{minipage}{0.40\textwidth}
  Experimental quantities are the probabilities to measure the
  observable $\Lambda$ in the state $\rho$
\end{minipage}\hfill
\begin{minipage}{0.22\textwidth}
\emph{they are calculated in theory as Born Probability}
\end{minipage}\hfill
\begin{minipage}{0.27\textwidth}
measured as ratio of many detector counts
\end{minipage}\\[5pt]
\begin{minipage}{0.19\textwidth}
$\mathcal{P}_{\rho}(\Lambda(t))\equiv$
\end{minipage}\hfill
\begin{minipage}{0.52\textwidth}
\[\Tr(\Lambda(t)\,\rho_{0})=\Tr(\Lambda_{0}\,\rho(t))\]
\end{minipage}\hfill
\begin{minipage}{0.14\textwidth}
$\approx N(t)/N$
\end{minipage}
\begin{minipage}{0.05\textwidth}
\flushright
\refstepcounter{equation}(\theequation)\label{eq:measrat}
\end{minipage}\\
\begin{equation}
\vert\langle\psi^{-}(t)\vert\phi^{+}\rangle\vert^{2}=
\vert\langle\psi^{-}\vert\phi^{+}(t)\rangle\vert^{2}
\end{equation}
\[ \text{in Heisenberg picture}\quad\quad\quad\quad
\text{in Schr\"odinger picture}
\]
The comparison between the theoretically calculated probability 
and the experimentally measured detector counts is given by
\begin{equation}
\mathcal{P}_\Lambda(\rho(t))\approx\frac{N(t)}{N} = \textrm{ ratio of registered detector counts},
\end{equation}
where $N$ is preferably a large number.

But in the theoretical description one identifies states with observables
by choosing the Hilbert Space Axiom:
\begin{equation}
\label{eq:hilbertaxiom}
\begin{array}{r}
\textrm{set of states }\{\phi\}=\mathcal{H}=\textrm{Hilbert space}\\
\textrm{and set of observables }\{\psi\}=\mathcal{H}=\textrm{Hilbert space}
\end{array}
\end{equation}
as boundary conditions for the dynamical differential equations:\\[5pt]
\begin{minipage}{0.45\textwidth}
\center the Heisenberg equation
\end{minipage}\hfill or \hfill
\begin{minipage}{0.45\textwidth}
\center the Schr\"odinger equation
\end{minipage}\\
\begin{minipage}{0.45\textwidth}
\begin{equation}
\label{eq:vecdynh}
i\hbar\frac{d}{d t}\psi(t)=-H\psi(t)
\end{equation}
\end{minipage} \hfill
\begin{minipage}{0.45\textwidth}
\begin{equation}
\label{eq:vecdyn}
i\hbar\frac{d}{d t}\phi(t)=H\phi(t)
\end{equation}
\end{minipage}\\[3pt]
\begin{minipage}{0.45\textwidth}
with state $\rho=\ketbra{\phi}{\phi}$ kept fixed
\end{minipage}  \hfill
\begin{minipage}{0.45\textwidth}
with observable $\Lambda=\ketbra{\psi}{\psi}$\\ kept fixed
\end{minipage}\\[3pt]
As a consequence of the Hilbert space boundary condition~\eqref{eq:hilbertaxiom} it follows
(Stone-von Neumann Theorem)
that all solutions of the differential equations are given by
the unitary group evolution~\cite{stone_1932,von_neumann_1932}:\\
\begin{minipage}{0.45\textwidth}
\begin{equation}
\label{eq:a}
\psi(t)=\text{e}^{iHt/\hbar}\psi,\quad-\infty<t<+\infty
\end{equation}
\end{minipage} \hfill or \hfill
\begin{minipage}{0.45\textwidth}
\begin{equation}
\label{eq:b}
\phi(t)=\text{e}^{-iHt/\hbar}\phi,\quad-\infty<t<+\infty
\end{equation}
\end{minipage}\\
\begin{minipage}{0.42\textwidth}
\emph{in the Heisenberg picture}
\end{minipage}\hfill
\begin{minipage}{0.42\textwidth}
\emph{in the Schr\"odinger picture}
\end{minipage}\\
and for the operators\\
\begin{minipage}{0.45\textwidth}
\begin{equation}
\label{eq:c}
\Lambda(t)=e^{i H t/\hbar}\Lambda(0)e^{-i H t/\hbar}
\end{equation}
\end{minipage}\hfill or \hfill
\begin{minipage}{0.45\textwidth}
\begin{equation}
\label{eq:d}
\rho(t)=e^{-i H t/\hbar}\rho(0)e^{i H t/\hbar}
\end{equation}
\end{minipage}\\
The time evolution extends from $t\rightarrow -\infty$ to $t\rightarrow +\infty$
and is given by
the unitary group $U(t)=e^{i H t/\hbar}$.
Every $U(t)$ has an inverse $U^{-1}(t)=U(-t)=U^\dagger(t)$.

Other boundary conditions for the same differential equations~\eqref{eq:vecdynh} and \eqref{eq:vecdyn}
result in different solutions of the Heisenberg/Schr\"odinger equation,
e.g. in $\psi(t)=\mathcal{U}(t-t_0)\psi(t_0)$
with $\mathcal{U}(t-t_0)=e^{i H (t-t_0)}$, $t_0=$ finite, e.g. $t_0=0$
(semigroup).
One is therefore justified to ask the question:
What is the experimental evidence for $t$ to be
$-\infty < t < +\infty$, i.e. for $U(t)$ to be a
\emph{group?}

In analogy with the radiation arrow of time, it is clear that an experimental
system must be prepared in a state, $\phi(t)$, before the observable $\ketbra{\psi}{\psi}$
can be measured in it (causality.)
For example, the detector cannot count decay products 
before the decaying state has been prepared.
This leads to a Quantum Mechanical Arrow of Time (QMAT):
The Born probability to measure the observable $\ketbra{\psi}{\psi}$
in the state $\phi(t)$,
\begin{equation}
\label{eq:qmat}
\mathcal{P}_\psi(\phi(t))=\vert \braket{\psi}{\phi(t)} \vert^2
=\vert \braket{\psi}{e^{-i H t/\hbar}\phi} \vert^2
=\vert \braket{e^{i H t/\hbar}\psi}{\phi} \vert^2
=\vert \braket{\psi(t)}{\phi} \vert^2,
\end{equation}
exists (experimentally) only for $t\geq t_0(=0)$,
where $t_0=0$ is the time that a system is prepared to be in the state
represented by $\phi$.

This is in contrast with the unitary group evolution~\eqref{eq:a}-\eqref{eq:d}
of conventional quantum mechanics, which predicts
$\vert \braket{\psi}{\phi(t)} \vert^2$ for all $-\infty < t < +\infty$.
Therefore one should have been aware that something might be wrong
with the unitary group evolution~\eqref{eq:a}-\eqref{eq:d}, and that 
a new theory is needed
for which the solutions of the Schr\"odinger equation, the states $\phi(t)$,
evolve by the \textit{semigroup}
\begin{eqnarray}
\label{eq:stsemigrp}
\phi^+(t)&=&\mathcal{U}^\times(t)\phi^+ \nonumber \\
\mathcal{U}^\times(t)&=&e^{-i H^\times t/\hbar}, \quad 0\leq t < \infty,
\end{eqnarray}
or for which the solutions of the Heisenberg equation, the observables $\psi(t)$,
evolve by the \textit{semigroup}
\begin{eqnarray}
\label{eq:semigroupobservables}
\psi^-(t)&=&\mathcal{U}(t)\psi^- \nonumber \\
\mathcal{U}(t)&=&e^{+i H t/\hbar}, \quad 0\leq t < \infty.
\end{eqnarray}
This means that we must mathematically distinguish between
the set of states (in-states of scattering experiments), $\{\phi\}$, and
the set of observables (detected out-states of scattering experiments), $\{\psi\}$.

\section{States and Observables in Scattering and Decay}
To make a mathematical distinction between the set of states and the set of observables
is suggested by the heuristic notions used in the
phenomenological description of scattering and decay:\\
1. One uses the Lippmann-Schwinger 
in- and out- plane wave ``states'' $\vert
E^{+}\rangle$ and $\vert E^{-}\rangle$, which fulfill the
Lippmann-Schwinger 
equation~\cite{lippmann_schwinger_scattering_theory_1950,gell-mann_goldberger_formal_theory_scattering_1953}:
\begin{equation}
\label{eq:LSeqn}
  \vert E^{\pm}\rangle=\vert E\pm i\epsilon\rangle =\vert E\rangle +
  \frac{1}{E-H\pm i\epsilon}V \vert E\rangle =\Omega^{\pm} \vert
  E\rangle,\quad \epsilon\rightarrow +0.
\end{equation}
\begin{minipage}{0.53\textwidth}
2. One continues the analytic $S$-matrix\\
to complex energies;
\end{minipage}\hfill
\begin{minipage}{0.42\textwidth}
\centerline{$S_{j}(E)\rightarrow S_{j}(z)$}
\end{minipage}\\
\begin{minipage}{0.53\textwidth}
3. For the Gamow states $\phi^G$\\
one integrates around the pole of $S_j(z)$
\end{minipage}\hfill
\begin{minipage}{0.42\textwidth}
\centerline{$z_{R}=E_{R}-i\Gamma/2$}
\centerline{(on the second sheet of $S$-matrix)}
\end{minipage}\\
\begin{minipage}{0.53\textwidth}
4. For the Lippmann-Schwinger equation\\
or in the propagator of field theory one uses
\end{minipage}\hfill
\begin{minipage}{0.42\textwidth}
\centerline{$z=E\pm i\epsilon,\quad \epsilon\mbox{ infinitesimal}$}
\end{minipage}\\
The kets $\vert E^+\rangle$ of~\eqref{eq:LSeqn} are taken as basis kets for a Dirac basis vector expansion of
in-state vectors (defined by the preparation apparatus of a scattering experiment)
\begin{equation}
\label{eq:epexp}
 \phi^{+}=\sum_{j,j_{3},\eta}\int_{0}^{\infty} dE \vert
  E,j,j_{3},\eta^{+} \rangle\langle^{+}
  E,j,j_{3},\eta\vert\phi^{+}\rangle
  = \int dE \ket{E^+}\braket{^+E}{\phi^+},
\end{equation}
and the $\ket{E^-}$ of~\eqref{eq:LSeqn} are taken for out-vectors
(representing observables $\ketbra{\psi^-}{\psi^-}$ defined by the detector)
of the scattering experiment
\begin{equation}
\label{eq:emexp}
 \psi^{-}=\sum_{j,j_{3},\eta}\int_{0}^{\infty} dE \vert
  E,j,j_{3},\eta^{-} \rangle\langle^{-}
  E,j,j_{3},\eta\vert\psi^{-}\rangle
  = \int dE \ket{E^-}\braket{^-E}{\psi^-}.
\end{equation}

The basis vector expansions like \eqref{eq:epexp}, \eqref{eq:emexp} were justified as
generalizations of more elementary formulas:
\begin{enumerate}
\item
The three components $x^i=\braket{i}{\vec{x}}$ define the vector in $\mathbb{R}^3$:
$\vec{x}=\sum_{i=1}^3\ket{i}x^i$.
\item
In analogy with this three-dimensional basis vector expansion, 
Dirac postulated that for a quantum system\footnote{
e.g. one with spherically symmetric Hamiltonian, $[H,J_i]=0$, where the $J_i$
are the angular momentum operators.
}
there exists a complete system of common eigenvectors $\vert Ejj_{3}\rangle$:
\begin{equation}
H\vert Ejj_{3}\rangle=E\vert Ejj_{3}\rangle,\quad
  J^{2}\vert Ejj_{3}\rangle=j(j+1)\vert Ejj_{3}\rangle,\quad J_{3}
  \vert Ejj_{3}\rangle=j_{3}\vert Ejj_{3}\rangle.
\end{equation}
The eigenvalues may be discrete and/or continuous, and every
solution of the Schr\"odinger or of the Heisenberg equation can be expanded as
\begin{equation}
\phi=\sum_{jj_{3}}\int _{0}^{\infty}dE\vert Ejj_{3}\rangle\langle Ejj_{3}\vert \phi\rangle.
\end{equation}

For continuous $E$, the $\vert Ejj_{3}\rangle$ are the Dirac kets.
Dirac kets are not in the Hilbert space
${\mathcal{H}}$, but they are new vectors that fulfill
the \textquotedblleft Dirac orthogonality condition\textquotedblright\
\begin{equation}
\langle E^{\prime}j^{\prime}j_{3}^{\prime}\vert Ejj_{3}\rangle=\delta(E^{\prime
}-E)\delta_{j^{\prime}j}\delta_{j_{3}^{\prime}j_{3}},
\end{equation}
where $\delta(E^\prime-E)$ is the generalization of $\delta_{E_{n^\prime}E_n}=\delta_{n^\prime n}$.
It took about 20 years to give a mathematical meaning to Dirac's $\delta(E^\prime-E)$.
\item
On the basis of Dirac's $\delta$-function, Schwartz created in 1945 the 
Theory of Distributions~\cite{schwartz_book_1950}.
In analogy with the Kronecker $\delta$, which fulfills
$\sum_{n^\prime}\delta_{E_{n^\prime}E_n}\braket{E_{n^\prime}}{\phi}=\braket{E_n}{\phi}$,
the distribution $\delta(E^\prime-E)$ is defined as the mathematical object
that fulfills the identity
\begin{equation}
\int dE^\prime\delta(E^\prime-E)\braket{E^\prime j j_3}{\phi}=\braket{E j j_3}{\phi}
\textrm{ or }
\int dE^\prime \delta(E^\prime-E)\phi(E^\prime)=\phi(E)
\end{equation}
for a space of ``well-behaved energy wave functions''\\
$\{\phi(E)\}=\{\braket{E j j_3}{\phi}\}=\{\phi_{j j_3}(E)\}$.
\end{enumerate}

In the Schwartz theory the Dirac delta, $\delta(E^\prime-E)$, is defined as an
antilinear functional on the ``space of well-behaved functions'' 
$\{\phi(E)\}\equiv\mathcal{S}$, the Schwartz space
(infinitely differentiable, rapidly decreasing)~\cite{schwartz_book_1950}.
The Schwartz space $\mathcal{S}$ fulfills
\begin{eqnarray}
\mathcal{S}\subset L^2[0,\infty). \nonumber
\end{eqnarray}
This means that 
\textbf{some} of the classes of Hilbert space functions $\{\phi_h(E)\}$ contain
also a smooth function $\phi(E)\in\mathcal{S}$.

Consider the space of continuous anti-linear functionals $\mathcal{S}^\times$ on
$\mathcal{S}$ and the continuous functionals $(L^2)^\times$ on $L^2$
(space of $L^2$-integrable functions.)
Then
\begin{eqnarray}
\label{eq:rhstriplet}
\{\phi(E)\} = \mathcal{S}\subset L^2 & = & (L^2)^\times\subset\mathcal{S}^\times
=\{\textrm{Distributions}\}.\\
& \uparrow & \nonumber
\end{eqnarray}
\vspace{-20pt}
\nopagebreak
\begin{center} because of the Fr\'echet-Riesz theorem \end{center}
This triplet~\eqref{eq:rhstriplet} is the Rigged Hilbert Space (RHS) of Schwartz space functions
$\mathcal{S}=\{\phi(E)\}$,
and the Dirac distribution is defined as a Schwartz space functional,
$\delta(E^\prime-E)\in\mathcal{S}^\times$.
The abstract Rigged Hilbert Space is the triplet of linear topological vector spaces,
\begin{equation}
\{\phi\}=\Phi\subset\mathcal{H}=\mathcal{H}^\times\subset\Phi^\times,
\end{equation}
that is (algebraically and topologically) isomorphic to~\eqref{eq:rhstriplet}.

The mathematical theory of Dirac kets 
uses \textbf{one} Schwartz space, $\Phi$.
This means it does \textbf{not} allow us to distinguish between 
prepared in-states $\{\phi^+\}$ (defined by the accelerator
of a scattering experiment)
and registered observables $\{\psi^-\}$ (defined by the detector
of a scattering experiment.)
This is in contrast to the Lippmann-Schwinger equations~\eqref{eq:LSeqn},
which suggest two different basis systems: one for the in-states, $\phi^+$,
and the other for the out-vectors, $\psi^-$.

As with the Hilbert space,
the dynamical differential equations~\eqref{eq:vecdynh}, \eqref{eq:vecdyn}
integrate under the \textbf{Schwartz space} boundary condition,
\begin{equation}
\label{eq:schspbc}
\phi\in\Phi=\textrm{ Schwartz space},\quad\psi\in\Phi=\textrm{ Schwartz space},
\end{equation}
to the group evolution with $-\infty < t < \infty$ as
in~\eqref{eq:a} and~\eqref{eq:b} (Proposition II Chapter IV of~\cite{bohm_monograph_1989}.)
It thus also disagrees with causality given by the QMAT~\eqref{eq:qmat},
which requires the semigroup~\eqref{eq:stsemigrp}, \eqref{eq:semigroupobservables}.

\section{Determining the Spaces of Prepared States, $\{\phi^+\}$, and Detected
Observables, $\{\psi^-\}$}
To determine the spaces of states and of observables,
we turn to the Lippmann-Schwinger equation~\eqref{eq:LSeqn}.
Because of the $i\epsilon$ in~\eqref{eq:LSeqn},
the energy wave functions in~\eqref{eq:epexp}, \eqref{eq:emexp} fulfill:
\begin{enumerate}
\item $\phi^+(E)=\braket{^+E}{\phi^+}$ are boundary values of analytic functions 
in the lower complex plane, second sheet of the $S$-matrix, and
\item $\psi^-(E)=\braket{^-E}{\psi^-}$ are analytic functions in the upper complex plane.
\end{enumerate}
The solutions of \emph{both} the Schr\"{o}dinger and the Heisenberg 
equations (state $\phi^+$  and observable $\psi^-$) have a
Dirac basis vector expansion, \eqref{eq:epexp} and \eqref{eq:emexp}.

We therefore want the continuous components, $\braket{^+E}{\phi^+}\equiv\phi^+(E)$,
of the prepared in-state, $\phi^+$, which fulfills the Schr\"odinger equation \eqref{eq:vecdyn},
to be rapidly decreasing (Schwartz space) functions that can be analytically continued
into the lower complex energy plane (second sheet of the $S$-matrix.)
And we want the components, $\braket{^-E}{\psi^-}\equiv\psi^-(E)$,
of the detected out-observable, $\psi^-$ (or $\ketbra{\psi^-}{\psi^-}$),
which fulfills the Heisenberg equation \eqref{eq:vecdynh}, to be rapidly decreasing
(Schwartz space) functions that can be analytically continued into the upper complex plane.
Then the complex conjugate, $\braket{\psi^-}{E^-}=\overline{\psi^-(E)}$,
can be analytically continued into the lower complex energy plane
(both in the second sheet of the $S$-matrix.)

To determine the properties of the energy wave functions
$\phi^+(E)=\braket{^+E}{\phi^+}$ 
and of the energy wave functions
$\psi^-(E)=\braket{^-E}{\psi^-}$,
we consider the probability amplitude $(\psi^-,\phi^+)$.
The $\vert (\psi^-,\phi^+) \vert^2$ is the probability to register the observable
$\ketbra{\psi^-}{\psi^-}$ defined by the detector, in the state $\phi^+$ prepared
by the accelerator.
Thus $\psi^-(t)$ evolves like an observable, as in~\eqref{eq:semigroupobservables}, and 
\textbf{not} like the out ``state'' of conventional scattering theory.
But like in conventional scattering theory, $(\psi^-,\phi^+)$ is expressed
in terms of the $S$-matrix element $S_j^{\eta^\prime\eta}(E)$ of 
angular momentum $j$.

This means that $(\psi^-,\phi^+)$ is given by
\begin{eqnarray}
  (\psi^{-},\phi^{+})
  &=&
  (\psi^{\text{out}},S\phi^{\text{in}}) \nonumber \\
  &=&\sum_{j}\int_{E_{0}}^{\infty} dE \sum_{j_{3}}\sum_{\eta,\eta'}
  \langle\psi^{-}\vert E,j,j_{3},\eta'^{-}\rangle S_{j}^{\eta'\eta}(E)
  \langle^{+}E,j,j_{3},\eta\vert\phi^{+}\rangle
\end{eqnarray}
or in simplified notation by
\begin{equation}
\label{eq:smatsimp}
  (\psi^{-},\phi^{+})=\int_{E_{0}}^{\infty} dE\langle\psi^{-}\vert E^{-}\rangle
  S_{j}(E)\langle^{+} E\vert\phi^{+}\rangle.
\end{equation}
Here the $j$-th partial $S$-matrix element $S_j^{\eta^\prime\eta}(E)$
comprises the dynamics of the scattering with angular momentum $j$.
In particular, resonance scattering and decaying states are associated with
the (first order) pole of $S_j^{\eta^\prime\eta}(E)$
in the lower complex energy plane on the second sheet.

Many physicists think that resonances are decaying states,
and a common belief is that
\begin{equation}
\frac{\hbar}{\Gamma}=\tau.
\end{equation}
There is ample evidence that all spontaneously decaying quantum systems
obey the exponential law $\mathcal{P}(t)\sim e^{-t/\tau}$~\cite{norman_tests_1988}.
Furthermore, the lifetime-width relation
$\tau=\hbar/\Gamma$ has been tested for the $3p^2\textrm{P}_{3/2}$ level of Na
to an accuracy that goes beyond the Weisskopf-Wigner approximation.
Both linewidth $\Gamma$~\cite{oates_linewidth_1996}
and lifetime $\tau$~\cite{volz_lifetime_1996}
have been measured with sufficiently high accuracy:\\
\begin{tabular}{ll}
the linewidth measurement gives & $\frac{\hbar}{\Gamma}=(16.237\pm 0.035)\,\textrm{ns}$ \\
the lifetime measurement gives & $\tau=(16.254\pm 0.022)\,\textrm{ns}$
\end{tabular}

\begin{figure}
\includegraphics[width=0.65\textwidth]{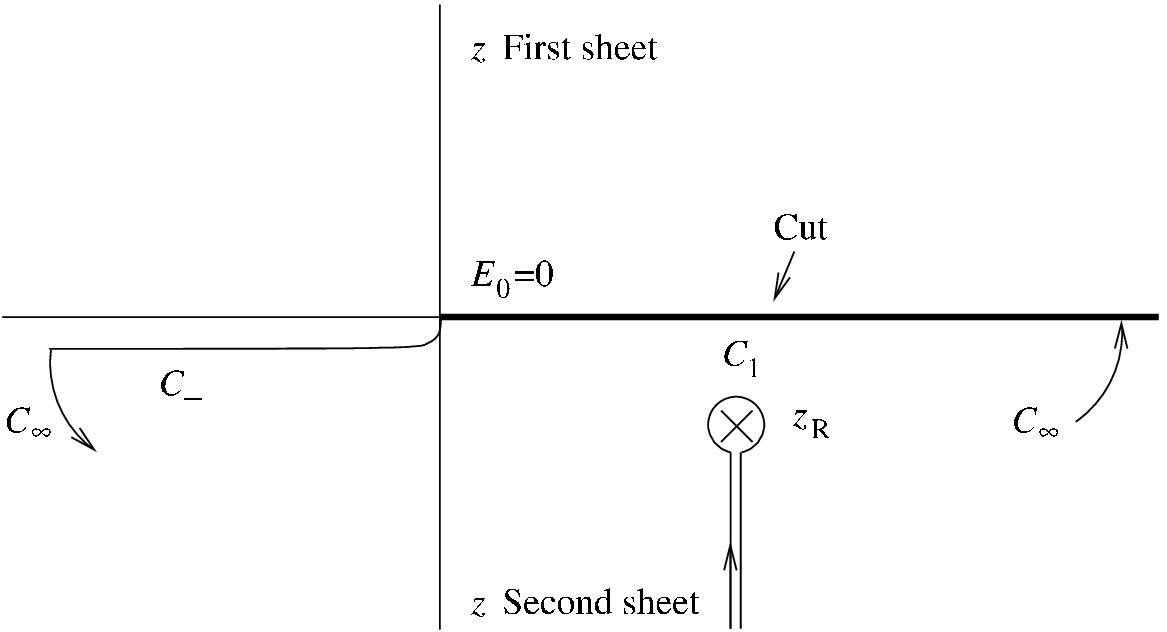}
\caption{Pole of the $S$-matrix.}\label{fig:pole}
\end{figure}

Therefore the program to determine the mathematical properties of the energy wave functions
$\braket{\psi^-}{E^-}$ and $\braket{^+E}{\phi^+}$ is:
Start with the pole term of the $S$-matrix element ($S_j(E)$ in~\eqref{eq:smatsimp}) at $z_{R}$,
as seen in Figure~\ref{fig:pole}, and determine the
mathematical property of $\langle^{-}E\vert\psi^{-}\rangle$ and
$\langle^{+}E\vert\phi^{+}\rangle$ in~\eqref{eq:smatsimp}
such that a Breit-Wigner resonance amplitude~\eqref{eq:bwamp}
as well as a decaying state vector~\eqref{eq:gamowket}
are derived from this $S$-matrix pole at $z_R$.

Then the Breit-Wigner resonance amplitude
\begin{equation}
\label{eq:bwamp}
a_j^{BW_i}=\frac{R_i}{E-z_{R_i}}
\end{equation}
for $z_{R_i}= E_{R_i}-i\Gamma_i/2$ is uniquely related to
a ket (functional)
\begin{equation}
\label{eq:gamowket}
\phi_j^G = \ket{z_{R_i},j,j_3,\eta^-}\sqrt{2\pi \Gamma_i}=\int dE\, 
\ket{E,j,j_3,\eta^-}\frac{i\sqrt{\frac{\Gamma}{2\pi}}}{E-z_R}
\end{equation}
with Breit-Wigner energy wave function
$\braket{E}{\phi^G}\sim\frac{1}{E-z_R}$.
Also, the ket $\phi^G$ should be a generalized eigenvector
with a discrete complex eigenvalue (as Gamow wanted),
\begin{equation}
H^\times \ket{E_R-i\Gamma/2^-}= (E_R-i\Gamma/2)\,\ket{E_R-i\Gamma/2^-},
\end{equation}
and
the ket $\phi^G$ should have exactly exponential time evolution,
i.e. it should fulfill
\begin{eqnarray}
\braket{\mathcal{U}(t)\psi_\eta^-}{\phi_j^G}
&=&\braket{\psi^-}{\mathcal{U}^\times(t)\phi^G_j}\sim \nonumber \\
\braket{e^{iHt/\hbar}\psi_\eta^-}{E_R-i\Gamma/2^-} 
&=& \bra{\psi_\eta^-}e^{-iH^\times t/\hbar}\ket{E_R-i\Gamma/2^-}\\
&=& e^{-iE_R t/\hbar} e^{-(\Gamma/2)t/\hbar} 
\braket{\psi_\eta^-}{E_R-i\Gamma/2^-}
\nonumber
\end{eqnarray}
\noindent
Then $\vert\braket{\psi^-(t)}{\phi^G}\vert^2=
e^{-\Gamma t/\hbar}\vert\braket{\psi^-(0)}{\phi^G}\vert^2$,
which means that a (Resonance state of width $\Gamma$) is
precisely a (Decaying state with lifetime $\tau=\frac{\hbar}{\Gamma}$),
and we have a theory that unifies resonance and decay phenomena.

The mathematical properties that one had to assume for the wave functions,
$\phi^+(E)=\braket{^+E}{\phi^+}$ and
$\psi^-(E)=\braket{^-E}{\psi^-}$ $\big(\braket{\psi^-}{E^-}=\overline{\psi^-(E)}\big)$,
were identified by H. Baumgartel (1977) as those of Hardy functions:
The energy wave functions of a prepared in-state, $\phi^+$, are Hardy functions analytic on the
lower complex plane, $\mathbb{C_-}$ (second sheet of the $S$-matrix),
\begin{equation}
\label{eq:hardyst}
\phi^+(E)=\braket{^+E}{\phi^+}\textrm{ Hardy on }\mathbb{C}_-.
\end{equation}
The energy wave functions of a detected out-state, $\psi^-$, or of
an observable, $\ketbra{\psi^-}{\psi^-}$, are Hardy functions
analytic on the upper complex plane, $\mathbb{C}_+$,
\begin{equation}
\label{eq:hardyob}
\psi^-(E)=\braket{^-E}{\psi^-}\textrm{ Hardy on }\mathbb{C}_+.
\end{equation}

Based on these conjectures, Manolo Gadella constructed
the two Gadella diagrams that provide the new axiom for the quantum theory of
scattering and decay.\\

\begin{minipage}{0.9\textwidth}
\centerline{The \textbf{Gadella Diagram} for observables is}
\begin{center}
\begin{tabular}{ccccc}
$\psi^-\in\Phi_+$    &  $\subset$   &  $\mathcal{H}$  &  $\subset$   &  $(\Phi_+)^\times$ \\[5pt]
$U^+\,\downarrow\qquad$  &    &  $\qquad\downarrow\,U^+$  & & $\qquad\downarrow (U^+)^\times$ \\[5pt]
$(\mathcal{H}^2_+\cap \mathcal{S})\vert_{\mathbb{R}^+}$ & $\subset$ & $\mathcal{L}^2[0,\infty)$ & 
$\subset$ & $\Big((\mathcal{H}^2_+\cap\mathcal{S})\vert_{\mathbb{R}^+}\Big)^\times$\\[5pt]
$(\theta_+)^{-1}\downarrow\qquad\quad\,$ & & & & $\qquad\downarrow (\theta^\times_+)^{-1}$\\[5pt]
$\mathcal{H}_+^2\cap\mathcal{S}$ & $\subset$ & $\mathcal{H}^2_+$ & $\subset$ & 
$\Big(\mathcal{H}^2_+\cap\mathcal{S}\Big)^\times$
\end{tabular}
\end{center}
\end{minipage}\hfill
\begin{minipage}{0.05\textwidth}
\flushright
\refstepcounter{equation}(\theequation)\label{gd:obs}
\end{minipage}\\[3pt]

The observables representing the detector, $A$, $\ketbra{\psi^-}{\psi^-}$,
are continuous operators in $\Phi_+$, $\psi^-\in\Phi_+$.
The detected out-``states,'' $\psi^-$, of scattering experiments are
elements of the abstract Hardy space, $\Phi_+$.
The Lippmann-Schwinger kets are now defined as functionals
$\ket{E^-}\in\Phi_+^\times$.
The energy wave functions
$\braket{^-E}{\psi^-}\in(\mathcal{H}^2_+\cap \mathcal{S})\vert_{\mathbb{R}^+}$
describe the detector efficiency ($\vert\braket{^-E}{\psi^-}\vert^2$ is 
the energy resolution of the detector.)
They are Hardy classes, $\mathcal{H}^2_+$, intersected
with elements of the Schwartz space, $\mathcal{S}(-\infty,\infty)$, and then restricted to the
positive real energy axis.\footnote{
Though the axiom~\eqref{gd:obs}, \eqref{gd:states} may look much more complicated
than the Hilbert space axiom, the 
$\phi^+(E)=\braket{^+E}{\phi^+}\in(\mathcal{H}_-^2\cap\mathcal{S})\vert_{\mathbb{R}^+}$
and
$\psi^-(E)=\braket{^-E}{\psi^-}\in(\mathcal{H}^2_+\cap \mathcal{S})\vert_{\mathbb{R}^+}$
are much nicer functions (smooth, rapidly decreasing, analytic) than the Lebesgue square integrable
functions of von Neumann's Hilbert space, $L^2[0,\infty)$.
}\\

\begin{minipage}{0.9\textwidth}
\centerline{The \textbf{Gadella Diagram} for states is}
\begin{center}
\begin{tabular}{ccccc}
$\phi^+\in\Phi_-$    &  $\subset$   &  $\mathcal{H}$  &  $\subset$   &  $(\Phi_-)^\times$ \\[5pt]
$U^-\,\downarrow\qquad$  &    &  $\qquad\downarrow\,U^-$  & & $\qquad\downarrow (U^-)^\times$ \\[5pt]
$(\mathcal{H}^2_-\cap \mathcal{S})\vert_{\mathbb{R}^+}$ & $\subset$ & $\mathcal{L}^2[0,\infty)$ & 
$\subset$ & $\Big((\mathcal{H}^2_-\cap\mathcal{S})\vert_{\mathbb{R}^+}\Big)^\times$\\[5pt]
$(\theta_-)^{-1}\downarrow\qquad\quad\,$ & & & & $\qquad\downarrow (\theta^\times_-)^{-1}$\\[5pt]
$\mathcal{H}_-^2\cap\mathcal{S}$ & $\subset$ & $\mathcal{H}^2_-$ & $\subset$ & 
$\Big(\mathcal{H}^2_-\cap\mathcal{S}\Big)^\times$
\end{tabular}
\end{center}
\end{minipage}\hfill
\begin{minipage}{0.05\textwidth}
\flushright
\refstepcounter{equation}(\theequation)\label{gd:states}
\end{minipage}\\[3pt]

The states (representing the preparation apparatus, e.g. accelerator: $\rho$, $\phi^+$)
are in $\Phi_-$.
The Lippmann-Schwinger kets are now mathematically defined as
$\ket{E^+}\in(\Phi_-)^\times$.
The energy wave functions
$\braket{^+E}{\phi^+}\in(\mathcal{H}_-^2\cap\mathcal{S})\vert_{\mathbb{R}^+}$
describe the energy distribution of the accelerator beam.\footnote{
In order to appreciate the mathematical importance of the 
Gadella diagrams, it should be mentioned that 
$(\mathcal{H}^2_\mp\cap \mathcal{S})\vert_{\mathbb{R}^+}$
means to intersect first $\mathcal{H}^2_\mp$ with the
Schwartz space $\mathcal{S}(-\infty,\infty)$
on the real line, $\mathbb{R}$, and then restrict the
intersection to the positive real line, $\mathbb{R}^+$.
The Breit-Wigner function in~\eqref{eq:gamowket} is an element of
$L^2(\mathbb{R})=\mathcal{H}^2_+\oplus\mathcal{H}^2_-$,
but it is a complicated distribution on the positive real axis,
$\mathbb{R}^+$~\cite{antoniou_complex_delta_1999}.
The multiplication operator on $\mathcal{H}^2_-$ has deficiency indices (0,1),
and on $\mathcal{H}^2_+$ it has deficiency indices (1,0).
On $(\mathcal{H}^2_\mp\cap \mathcal{S})\vert_{\mathbb{R}^+}$, however,
it has deficiency indices (0,0).
Thus the Hamiltonian is essentially self-adjoint in $\Phi_+$ 
and in $\Phi_-$.
}

Replacing the Hilbert space axiom~\eqref{eq:hilbertaxiom}
or the Schwartz space axiom~\eqref{eq:schspbc}
with the axiom given by
the Gadella Diagrams~\eqref{gd:obs}, \eqref{gd:states},
one obtains kets $\ket{z^\pm}$, $\ket{E^\pm}$, $\ket{E_R-i\Gamma/2^-}$ that are now
mathematically well defined as functionals on the spaces $\Phi_\mp$:
$\ket{E^\pm}\in\Phi^\times_\mp$.
One has finally a consistent mathematical theory that unifies resonance
and decay phenomena, with $\Gamma=\frac{\hbar}{\tau}$ and with an exact
exponential decay law.

But in place of the unitary group evolution as a consequence of the
Hilbert space or Schwartz space axiom~\eqref{eq:hilbertaxiom},
\eqref{eq:schspbc}, the solutions of the dynamical equations are now to be solved
under the Hardy space boundary conditions.
For that one has in place of the Stone-von Neumann theorem
the Paley-Wiener theorem~\cite{paley_wiener_1934}, from which it follows that
the time evolution is
given by the two \textit{semigroups}:
\begin{equation}
\label{eq:semigrpobs}
\psi^-(t)=e^{i H (t-t_0)/\hbar}\psi^-\quad t_0\leq t<\infty \quad
\end{equation}
for the observables (Heisenberg equation), and 
\begin{equation}
\label{eq:semigrpstate}
\phi^+(t)=e^{-i H (t-t_0)/\hbar}\phi^+\quad t_0\leq t<\infty
\end{equation}
for the states (Schr\"odinger equation.)
A straightforward consequence of the semigroup solution
of the dynamical equations is that~\eqref{eq:semigrpobs}, \eqref{eq:semigrpstate}
predict the Born probabilities
\begin{equation}
\label{eq:semibrn}
\vert\braket{\psi^-(t)}{\phi^+}\vert^2 \quad \textrm{only for } t\geq t_0.
\end{equation}
This expresses causality:

The probability to find the observable
$\ketbra{\psi^-(t)}{\psi^-(t)}$ in the state $\phi^+$ is predicted only
for a time $t$ after the time $t_0$ at which the state of the experimental
system had been prepared.\footnote{
This is the analogue of the radiation arrow of time (Sommerfeld
radiation condition): A source (transmitter) must emit radiation
(at $t_0$) before the radiation can be detected by a receiver
at $t > t_0$.
}

It also predicts---in contrast to the group evolution~\eqref{eq:a}, \eqref{eq:b}
for the Hilbert space axiom and for the Schwartz space axiom---a
beginning of time $t=t_0$ for quantum states,
as it is required by the QMAT~\eqref{eq:qmat}.

When in an era dominated by group theory and Hilbert space axiomatics,
one derives from the axioms that unify Breit-Wigner resonances and exponential decay,
the quantum mechanical time asymmetry~\eqref{eq:semigrpobs}-\eqref{eq:semibrn},
one has a shocking result.
The QMAT~\eqref{eq:qmat} provides intuitive support for this result,
and there exists an analogy with the well-accepted radiation arrow of time, but one
still needs to ask the question:
How can one observe this beginning of time, $t_0$, in experimental data?

\section{Quantum jumps and the beginning of time}
\label{sec:qjumps}
Because quantum theory makes probabilistic predictions, all useful experiments in
quantum physics are performed on large ensembles of physical systems (such as
elementary particles, atoms, or ions.)
In the past, an ensemble has usually contained a large
number of ensemble members present at a given time in the laboratory,
and ensemble members have not been observed individually.
When many members are present simultaneously, there is no way to distinguish
by a clock in the laboratory which ensemble member has been prepared at a time, say $t_0^{(1)} $, and
which one at a time $t_0^{(2)} $, etc.

Recently, however, it has become possible to excite a \textit{single}
ion~\cite{nagourney_dehmelt_shelved_1986,yu_nag_deh_prl_1997,sauter_toschek_qjumps_1986,bergquist_qjumps_1986,peik_qjumps_1994}
into a metastable state $|m\rangle$ at a
time $t_0^{(1)}$, which can be measured precisely.
One repeats the preparation process $(M-1)$ number of times at
$t_0^{(2)},t_0^{(3)},\cdots t_0^{(M)}\equiv\{t_0^{(i)}\}$,
and then one has an ensemble of $M$ individual ions, identically prepared.
This ensemble of $M$ single ions is the experimental ensemble, and it is not
to be confused with a many-particle state.
Ignoring the possible degeneracies of the level $\ket{m}$, one represents
the state of every member of
the experimental ensemble by a state vector, $\phi(t)$.
Whereas usually one thinks of a quantum
mechanical ensemble as an ensemble of many members
present simultaneously in the lab, if one makes an experiment on
single ions, one has to prepare the single ion at an ensemble of
many times, $\{t_0^{(i)}\}$, and in an identical way, to obtain results described 
by quantum mechanics.

\begin{figure}
\includegraphics[trim = 4cm 20cm 3cm 5cm, clip=true, width=1\textwidth]{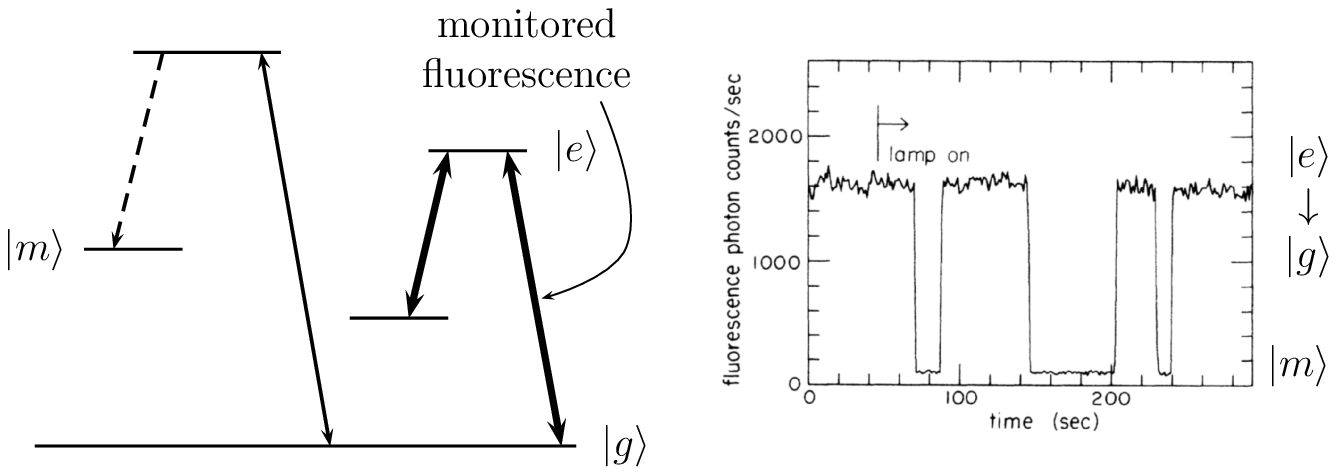}
\caption{Energy level diagram (left) and the
associated quantum jumps (right) copied from~\cite{nagourney_dehmelt_shelved_1986}.
On the left,
thick lines show laser-driven transitions; the thin line is a transition driven by
a lamp; and the dashed line represents a spontaneous decay into the metastable level,
$\ket{m}$.
When the ion is excited by the lamp and then decays to the level $\ket{m}$,
it cannot emit the monitored fluorescence.
On the right,
the sudden drop in fluorescence shown for three quantum jumps defines the
beginning of time for the $i$th member of the experimental ensemble.
The dark period's length on the time axis, $t^{(i)}$, is
the lifetime of the $i$th ensemble member.}
\label{fig2}
\end{figure}

Examples of such experiments on metastable states of single ions have been performed
in~\cite{nagourney_dehmelt_shelved_1986,yu_nag_deh_prl_1997,sauter_toschek_qjumps_1986,bergquist_qjumps_1986,peik_qjumps_1994}
using Dehmelt's idea~\cite{dehmelt_proposal_1975} of shelving the single ion on
a metastable level.  
Experiments of this kind require an ion with energy levels similar
to those shown on the left of Figure~\ref{fig2}.
All of the excited states, including the metastable state $|m\rangle$, are  radiatively
coupled to the same ground state $|g\rangle$, but $\ket{m}$ has a transition (decay) rate
to $\ket{g}$ vastly different from the transition rates of the others.

The thick lines indicate stimulated transitions driven by
lasers, and the intensity of the fluorescence from the transition
$\ket{e}\rightarrow\ket{g}$ is monitored.
Occasionally a lamp drives the transition shown with a thin line instead,
and the single ion lives in the highest excited state for a relatively short
time ($\approx 6\,\textrm{ns}$) before decaying to the metastable level $\ket{m}$.

While the ion is in the metastable level, it is ``shelved'' and cannot participate
in the $\ket{e}\leftrightarrow\ket{g}$ transition.
The duration of the shelf-time spent in the level $\ket{m}$ is therefore observed as
a dark period in the fluorescence (three examples are shown in Fig.~\ref{fig2}.)
On average, the dark periods in~\cite{nagourney_dehmelt_shelved_1986}
have a duration of about $30\,\textrm{s}$.

On the right of Figure \ref{fig2} is a typical experimental fluorescence measurement as a function
of time in the laboratory~\cite{nagourney_dehmelt_shelved_1986}.
We see a sudden onset of a
period of no fluorescence at times $t_0^{(i)}$, followed by a sudden return
of the original fluorescence intensity at later times $t_1^{(i)}$, with
$i=1,2,\cdots M$.  In the experiment of~\cite{nagourney_dehmelt_shelved_1986},
there were $M=203$ such dark periods; three of these are shown 
on the right of Figure~\ref{fig2}.

The onset of a dark period at $t_0^{(i)}$ indicates that the single ion
has been shelved in the metastable state $|m\rangle$.
The ion ``lives'' there for the duration of the length of the dark period
\begin{equation}
\label{vd}
t^{(i)} = t_1^{(i)} - t_0^{(i)}\,,
\quad i=1,2,\ldots M,
\end{equation}
because the return of fluorescence $|e\rangle\rightarrow|g\rangle$
at the later time $t_1^{(i)}$ indicates that the single ion is no
longer in the level $|m\rangle $. Each individual metastable ion
$|m\rangle $ is prepared at the time $ t^{(i)}_0$, and then transitions
at the time $t_1^{(i)}$ to the ground state $|g\rangle$.
The ion decays from the state $|m\rangle$ with the emission of a photon.
This single photon is not detectable, but $t_1^{(i)}$ is observed as the
onset time of subsequent florescence transitions.
One experimentally determines the duration of the dark period $t^{(i)}$ of \eqref{vd}
by reading from the time axis in Figure~\ref{fig2} the length of the $i$th dark period.

The survival probability of an ion in a metastable state, $\ket{m}$, is the probability
that, at a given duration in time from when it was initially prepared to be in $\ket{m}$,
the ion remains in $\ket{m}$ (that it has not decayed.)
From the ensemble of dwell times, $\{t^{(i)}\}$,
one determines the experimental survival probability 
as a function of the duration in time from when the ion system was initially prepared.
We denote by
\begin{equation}
\label{eq:numdef}
N_{|m\rangle}(t)\equiv N_{|m\rangle}(t:\,t^{(i)} > t)
\end{equation}
the number of dwell times of duration \emph{longer} than $t$.

\begin{figure}
\includegraphics[width=0.5\textwidth]{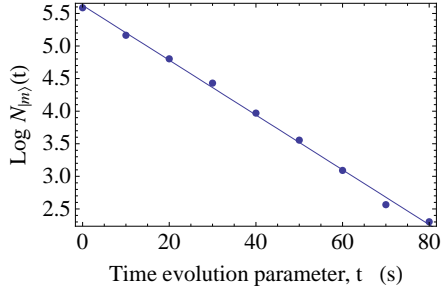}
\caption{Logarithmic plot of the experimentally
         determined number of dwell times of duration greater than the time parameter, $t$.
         It is created from results in~\cite{nagourney_dehmelt_shelved_1986},
         but we have not calculated error bars.
         This is not a survival probability because it is not normalized.
         It does, however, demonstrate the exponential character of decay, and
         the lifetime, $\tau$, is equal to
         the negative inverse of the slope of the line overlaid on the data.}
       \label{figdecaylaw}
\end{figure}

Figure~\ref{figdecaylaw}, which was created from the
histogram published in~\cite{nagourney_dehmelt_shelved_1986},
is a plot of $N_{\ket{m}}(t)$
for an ion prepared initially in the metastable level $|m\rangle$\footnote{
Because of the finite number of dwell times measured in the experiment,
$N_{|m\rangle}(t)$ in Figure~\ref{figdecaylaw} was only determined
for the time parameter in ten-second increments:
$t=0,10,20,\ldots\,\textrm{s}$~\cite{nagourney_dehmelt_shelved_1986}.
}.
The $t$ in~\eqref{eq:numdef} and in Figure~\ref{figdecaylaw} is 
not to be confused with the time (coordinate) marked by clocks in the lab 
(horizontal axis of Figure~\ref{fig2}.)
Rather, it is the time evolution parameter to which the \emph{duration}
of each dwell time is compared.
An experimental duration is always positive, so the $t$
in~\eqref{eq:numdef} always satisfies $t\geq 0$.
This is a manifestation of the quantum mechanical arrow of time~\eqref{eq:qmat}.

When properly normalized, Figure~\ref{figdecaylaw} is the experimental survival probability.
It can be compared to the theoretical survival probability,
which is calculated in the form of a Born probability~\eqref{eq:qmat}.
For the single ion experiments, this
is the probability that an ion in the state $\phi^+(t)$ is measured
still to be on the metastable level $|m\rangle$
at the time parameter value $t$:
\begin{equation}
\label{eq:born_prob_m}
\mathcal{P}_{|m\rangle}\big(\phi^+(t)\big)=\vert \langle m| \phi^+(t)\rangle\vert^2.
\end{equation}
Explicitly, the comparison between theory and experiment is the comparison
\begin{equation}
\textrm{counting ratio}\equiv
\frac{N_{|m\rangle}(t)}{M}\stackrel{?}{=}\mathcal{P}_{|m\rangle}\big(\phi^+(t)\big)
\equiv\textrm{ Born probability},
\label{eq:comparison}
\end{equation}
where $M$ is the total number of dwell times (dark periods) measured in the experiment.
Because the experimental durations, numbering $N_{\ket{m}}(t)$,
on the left hand side of~\eqref{eq:comparison}
are compared to a time parameter satisfying $t\geq 0$, the time evolution 
parameter of the state vector, $\phi^+(t)$, on the right hand side also satisfies $t\geq 0$.

Given~\eqref{eq:comparison}, one identifies
the beginning time, $t_0$ of~\eqref{eq:semibrn}, and thus of the semigroup~\eqref{eq:semigrpstate},
with the ensemble of onset times of the dark periods.
One has an ensemble of single ions, each
prepared to be in a metastable state $|m\rangle $ at an ensemble of
M preparation times, $\{t^{(i)}_0\}$, measured by clocks in the laboratory.
To this ensemble of times in the laboratory corresponds the value $t=0$
of the time parameter, $t$,  that parametrizes the evolution of the state vector,
$\phi^+(t)$:
\begin{equation}
\{ t_0^{(i)}:\, i=1,2,\ldots,M \} \Leftrightarrow t=0.
\end{equation}


\bibliographystyle{plain}       
\bibliography{taqmqj}   

\end{document}